\pdfoutput=1

\documentclass[a4paper,floatfix,pdftex,superscriptaddress,twoside,%
aip,rsi,reprint, final,
citeautoscript, 
]{revtex4-2}%
\usepackage{amsfonts,amsmath,amssymb}
\usepackage{commath}
\usepackage[T1]{fontenc}
\usepackage{graphicx}%
\usepackage[utf8]{inputenc}
\usepackage{microtype}
\usepackage{xspace}
\usepackage{hyperref, hypernat}
\usepackage{color} 
\usepackage{siunitx}
\usepackage[displaymath,textmath,graphics]{preview}
\usepackage[version=3]{mhchem}

\graphicspath{{./Fig/}} 

\newcommand*{\um}{\ensuremath{\text{\textmu{m}}}\xspace}%
\newcommand{\ude}{\affiliation{Faculty of Physics and Center of Nanointegration (CENIDE), University of Duisburg-Essen, Lotharstr. 1, 47057 Duisburg, Germany}}%

\begin{document}
\title{Integration of Conventional Surface Science Techniques with Surface-Sensitive Azimuthal and Polarization Dependent Femtosecond-Resolved Sum Frequency Generation Spectroscopy}
\author{\mbox{Zhipeng Huang}}
\email[Email:~]{zhipeng.huang@uni-due.de}\ude
\author{Tobias Roos}\ude
\author{\mbox{Yujin Tong}}\ude%
\author{\mbox{R.\ Kramer Campen}}
\email[Email:~]{richard.campen@uni-due.de}%
\ude
\date{\today}%

\begin{abstract}\noindent%
Experimental insight into the elementary processes underlying charge transfer across interfaces has blossomed with the wide-spread availability of ultra-high vacuum set-ups that allow the preparation and characterization of solid surfaces with well-defined molecular adsorbates over a wide ranges of temperatures. 
Within the last 15 years such insights have extended to charge transfer heterostructures containing solids overlain by one, or more, atomically thin two dimensional materials. 
Such systems are of wide potential interest \emph{both} because they appear to offer a path to separate surface reactivity from bulk chemical properties \emph{and} because some offer completely novel physics, unrealizable in bulk three dimensional solids. 
Thick layers of molecular adsorbates or heterostructures of 2D materials generally preclude the use of electrons or atoms as probes. 
However with linear photon-in/photon-out techniques it is often challenging to assign the observed optical response to a particular portion of the interface.
We and prior workers have demonstrated that by full characterization of the symmetry of the second order nonlinear optical susceptibility, \textit{i.e.}\ the $\chi^{(2)}$, in sum frequency generation (SFG) spectroscopy, this problem can be overcome.
Here we describe an ultra-high vacuum system built to allow conventional UHV sample preparation and characterization, femtosecond and polarization resolved SFG spectroscopy, the azimuthal sample rotation necessary to fully describe $\chi^{(2)}$ symmetry and with sufficient stability to allow scanning SFG microscopy. 
We demonstrate these capabilities in proof-of-principle measurements on CO adsorbed on Pt(111) and of the clean Ag(111) surface. 
Because this set-up allows \emph{both} full characterization of the nonlinear susceptibility \emph{and} the temperature control and sample preparation/characterization of conventional UHV set-ups we expect it to be of great utility in investigation of both the basic physics and applications of solid, 2D material heterostructures. 
\end{abstract}
\pacs{}
\maketitle

\section{Introduction}
\label{sec:introduction}

Understanding the properties of solid surfaces and their molecular adsorbates is important for designing better catalysts, developing new materials and in understanding elemental cycling in the environment~\cite{Bare:Intr:2003,ban23a}. 
Much physical insight into such systems has been gained over the last sixty years by the study of well defined single crystalline solids in contact with small numbers of adsorbed molecules in ultra-high vacuum (UHV).
Under UHV conditions surfaces are now routinely characterized using electron-based (\textit{e.g.}\ Low Energy Electron Diffraction (LEED)~\cite{Hove:LEED:1986}, Auger Electron Spectroscopy (AES)~\cite{Chang:AES:1971} and Electron Energy Loss Spectroscopy\cite{hof16}), atom-based (\textit{e.g.}\ Temperature Programmed Desorption Spectroscopy (TPD)~\cite{King:TPD:1975} and Helium atom scattering\cite{hol21}), photon-based (\textit{e.g.}\ Infrared Reflection Absorption Spectroscopy (IRAS)\cite{yan17,kes17}) and scanning probe techniques \cite{bot96}. 
Such tools offer direct insight into the morphology and electronic structure of clean solid surfaces.
Application to systems containing adsorbates further allows characterization of adsorbate structure and the thermodynamics and mechanism of ad(de)sorption.
With the increasing availability of pulsed photon and electron sources more recently structural fluctuations in such systems have been explored on timescales ranging from femto- to microseconds \cite{bac05,vog18,sto21}. 

Over the last several decades it has become clear that the properties of solid surfaces in contact with other condensed phases may differ dramatically from the same solids in contact with gasses \cite{sch15c}. 
However, because the mean free paths of electrons and atoms in condensed matter are $\lesssim$ nm such systems are generally only amenable to characterization by scanning probe or photon based techniques \cite{zae12}. 
Unfortunately, linear optical approaches, \textit{e.g.}\ infrared or x-ray absorption, are inherently bulk sensitive.
This sensitivity makes it challenging, for example, to extract the spectral response of molecular adsorbates at a solid surface in the presence of a 10$^{10}$ more of the same molecule in an adjoining bulk liquid phase. 
Even-order nonlinear optical techniques, \textit{e.g.}\ sum and difference frequency generation (SFG and DFG) spectroscopy, are interface-specific in the dipole approximation \cite{Shen:SFG:1989}.
Thus, these approaches, in their vibrationally or electronically resonant variants, are interface specific analogs of bulk sensitive IR or UV/Vis absorption \cite{du93,du94b,yam15b}.   
Because SFG/DFG spectroscopies are photon-in/photon-out, they are equally applicable to both solid/vacuum, solid/solid and solid/liquid interfaces \cite{Roiaz:operandoSFG:2018,Li:PSFG:2020}. 
Further, because they require intense ultrashort laser pulses, they lend themselves naturally to the characterization of ultrafast dynamics at interfaces between condensed phases \cite{Bonn:TSFG:2000}.   

SFG/DFG are interface-specific for systems with an interface between two bulk phases that are inversion symmetric and can be described in the dipole approximation \footnote{For practical considerations -- sensitive Si-based detector are widely available for light at visible wavelengths while detectors in for near-infrared wavelengths are generally less sensitive -- the great majority of reported second order nonlinear optical spectroscopy employs a sum (and not difference) frequency approach. To aid in concise exposition we therefore refer only to SFG going forward, although similar insights are, in principle, available from DFG}. 
For systems which meet this requirement, measurement of the SFG response offers insight into the orientation of molecular adsorbates that exceeds that available from polarization resolved infrared absorption (\textit{i.e.}\ multiple moments of adsorbate orientational distribution and applicability to systems with low infrared reflectivity) \cite{Zhuang:PSFG:1999,wei01}. 
However, even in systems that do not meet this restriction, \textit{e.g.}\ solids in which the bulk lattice lacks inversion symmetry or systems where the interface is charged, physical insight not available from linear optical spectroscopies is often possible.
To understand why, it's fruitful to remind ourselves that linear optical techniques probe the first order linear optical susceptibility: $\chi^ {(1)}$, a rank two tensor. 
SFG, in contrast, is a second order non-linear optical process in which the material response is given by the second order non-linear optical susceptibility: $\chi^{(2)}$, a rank three tensor \cite{Lambert:SFG:2005}.
Because many of the non-zero terms in the $\chi^{(2)}$ can be probed individually, by detecting the change in intensity of the emitted sum frequency field when changing the polarizations or angles of the the three fields or orientation of the sample, it is generally possible to experimentally characterize the symmetry of the spectral response \cite{Lambert:SFG:2005,Wang:SFG:2005} in a manner not possible in linear approaches.

Such characterization is important because for many systems of interest the symmetry of the optical response of material \textit{at} the interface differs from that in the adjoining bulk phase. 
For example, bulk $\alpha$-\ce{SiO2} belongs to the $D_{3}$ point group.
It is, therefore, non-inversion symmetric and thus bulk SFG active.  
This symmetry is, necessarily, decreased with any termination of the bulk lattice. 
As shown by Liu and Shen, these differences allow the isolation of the surface (optical) phonon response of $\alpha$-quartz(001), \textit{i.e.}\ surface metal-oxygen spectral response, even in the presence of the much larger contribution from bulk $\alpha$-\ce{SiO2} \cite{liu08,liu08b}.
Along similar lines, charged buried interfaces induce a field that propagates into, either, bulk condensed phase.
For example, at a semiconductor/liquid interface the intensity of the emitted sum frequency field is a function of \emph{both} the field gradient across the semiconductor's space charge layer \emph{and} the electric double layer extending into solution \cite{lan94,ong92,gei09}. 
This DC field breaks inversion symmetry over its characteristic screening length, in either phase, thus making all matter within this volume SFG active.
A variety of groups have shown that, given knowledge of the symmetry of the $\chi^{(2)}$, one can measure spectra associated with each portion of such interfaces.
For example, at a silica/water interface the spectrum of water hydrogen-bound to the nearby silica surface can be distinguished from those several nm away but within the adjoining electric double layer \cite{wen16a,ohn17}. 
Viewed more generally this body of work has clarified that for systems containing two bulk phases and an interface, multiple distinct, near-interface, regions may contribute to the emitted SFG and that these regions may be distinguished experimentally by fully characterizing the symmetry of a system's $\chi^{(2)}$. 

Within the last several decades much work has demonstrated that atomically thin two dimensional materials often have mechanical, optical or electrical properties that differ dramatically from bulk phases of the same stoichiometry \cite{zen18,sha22b,kum23}. 
While initial interest focussed on the properties of single, isolated monolayers, it has become increasingly clear that use of such materials in devices inevitably requires placing them in contact with strongly interacting bulk phases\cite{sie21} and that heterostructures composed of two (or more) atomically thin monolayers offer intriguing physics not present in isolated  monolayers or either bulk phase \cite{he21,beh21a}. 

Because coulomb screening is substantially reduced in monolayers relative to bulk phases of the same stoichiometry, semiconducting 2D materials, \textit{e.g.}\ Transition Metal Dichalcogenides (TMDCs), support excitons with dramatically increased binding energies relative to conventional semiconductors (several hundred vs tens meV). 
As a consequence their optical response is, for monolayers weakly interacting with substrates, dominated by excitons both in photoluminescence and absorption.
Placing TMDCs on strongly interacting, metal substrates, quenches the photoluminescence response of excitons and makes the optical response of the monolayer TMDC challenging to observe in reflection absorption measurements (the optical response of the TMDC is partially masked by that of free electrons in the metal) \cite{par18,pol21}. 

In addition to this linear optical characterization, there have been extensive nonlinear optical, principally second harmonic generation, characterization of monolayer 2D layered materials in general, and TMDCs in particular \cite{wan19f,zha20g}.
Taken as a whole this work suggests that monolayer 2D materials have nonlinear optical susceptibilities similar to bulk materials typically used in nonlinear photonics applications and therefore offer intriguing possibilities for devices in unconventional geometries. 
They also demonstrate that the symmetry of the second order nonlinear susceptibility, \textit{i.e.}\ the $\chi^{(2)}$, sensitively reports on strain, phase and monolayer orientation (with respect to the laboratory reference frame) \cite{li13d,kha20c}.
The intensity of the emitted second harmonic field increases by $> 1000 \times$ when the photon energy of either of the two incident 
fields or the emitted second harmonic emission is in resonance with an optically accessible transition \cite{Lambert:SFG:2005}. 
This suggests, as recently suggested by Zhumagulov et al.\ , that characterization of resonant $\chi^{(2)}$ symmetry should allow the quantitative description of exciton symmetry and its deviation from that of the lattice\cite{zhu22b}.

Second harmonic generation is an energetically degenerate version of sum frequency generation usually performed with both interactions coming from the same incident beam (and thus having the same angle with respect to the sample, the same photon energy and the same polarization). 
An SFG measurement with independent control over polarization, photon energy and incident angle of each of the two incident fields offers the maximum possible access to the different components of $\chi^{(2)}$. 
We have recently shown that a full polarization analysis of the SFG response of \ce{MoS2} adsorbed on Au allows quantitative separation of the optical response of the free electrons from Au or from adsorbed \ce{MoS2} \cite{Yang:ASFG:2023}. 
While the resulting sample shows no excitonic optical response, correct understanding of the symmetry of the SFG signal allows the separation of the optical contribution of the \ce{MoS2} related states from that of free electrons in Au and the quantification of substrate-induced bandgap renormalization in an all-optical configuration~\cite{yang2023isolating}. 
Characterizing charge carrier dynamics in such systems, \textit{e.g.}\ \ce{MoSe2}/\ce{WSe2} heterostructures \cite{riv16}, monolayer \ce{WS2} on Ag(111) \cite{uls17} or an indium tin oxide/\ce{MoS2}/aqueous NaI solution electrochemical cell, offers similar challenges to those described above \cite{aus23}. 
That is, one would like to be able to independently address each component of the interfacial system when all may contribute to the measured nonequilibrium optical response.

Four practical challenges characterize such studies. 
First, production of large, \textit{i.e.}\ > 1 mm, individual flakes of 2D materials is quite challenging.
As a result studies employing nonlinear optical techniques typically require an additional optical microscope to help locate the, focussed, laser spots on the sample. 
Sample stages which allow both identification of the laser spot location and precise translation would offer additionally the possibility of conducting scanning imaging on the \textmu m length scale over which flakes of 2D materials often vary. 
Second, the most straightforward way to probe the symmetry of the 2D-material's $\chi^{(2)}$ is to measure the change in emitted SFG field as a function of sample rotation around the surface normal.
Third insight into the basic physical processes that control charge carrier lifetime and relaxation mechanism (\textit{e.g.}\ the participation of phonons), typically requires conducting temperature dependent measurements over temperature ranges > 150 K.
As a practical matter such measurements require a vacuum cryostat or UHV chamber. 
Fourth if one additionally wishes to characterize, for example, the interaction of a 2D-material with a molecular organic phase, ultra-high vacuum conditions, and associated tools to characterize the deposited organic layer, are required. 

To our knowledge while multiple studies exist that measure polarization resolved, femtosecond time resolved and azimuthal angle dependent SFG response in a UHV chamber (with a large range of temperature control and ancillary sample creation and characterization tools available) there are no reports of a set-up that combines \emph{all} of these characteristics with the ability to image the sample with $\approx$ 50~\textmu m spatial resolution. 
Such a set-up is required if, for example, we are to fully understand the basic physics that control valley polarization in TMDCS as a function of substrate.  

Here we report on our newly built ultra-high vacuum setup, which combines \textit{i.e.}\ conventional UHV surface science sample preparation and characterization, femtosecond resolved sum frequency generation spectroscopy and a sample manipulator that allow high precision translation and rotation of the sample. 
In what follows we first demonstrate system capabilities similar to existing systems.
We do so by preparing a pristine Pt(111) surface -- demonstrated to be contaminant free by Auger Electron Spectrosopy (AES) and low defect density by comparison to published Low Energy Electron Diffraction (LEED) results. 
After preparation the surface is dosed with CO. 
CO orientation is characterized by polarization resolved vibrationally resonant sum frequency generation spectroscopy and CO adsorption energy by temperature programmed desorption. 
The ability to probe femtosecond resolved dynamics is demonstrated by collection of the free induction decay of the adsorbed CO. 
We next demonstrate that the system allows accurate collection of azimuthal dependent SFG, by collecting azimuthal angle dependent SFG from the Ag(111) surface and comparing the resulting signal's symmetry to that of the LEED pattern collected from the same sample. 
Finally we show, by comparing the scanned SFG image of a roughened Ag(111) crystal and a optical micrograph collected from the same sample, that our system is capable of collecting SFG images with $\approx$ 10~\textmu m spatial resolution. 
These proof of principle measurements show the system offers a novel combination of existing approaches that are of particular potential utility in the characterization of the properties of charge carriers and their relaxation in 2D-materials.

\section{Results and Discussion}
\label{sec:results}

\subsection{Description of our UHV setup}
\label{sec:UHV}
\begin{figure}[!htbp]
	\centering
	\includegraphics[width=0.49\textwidth]{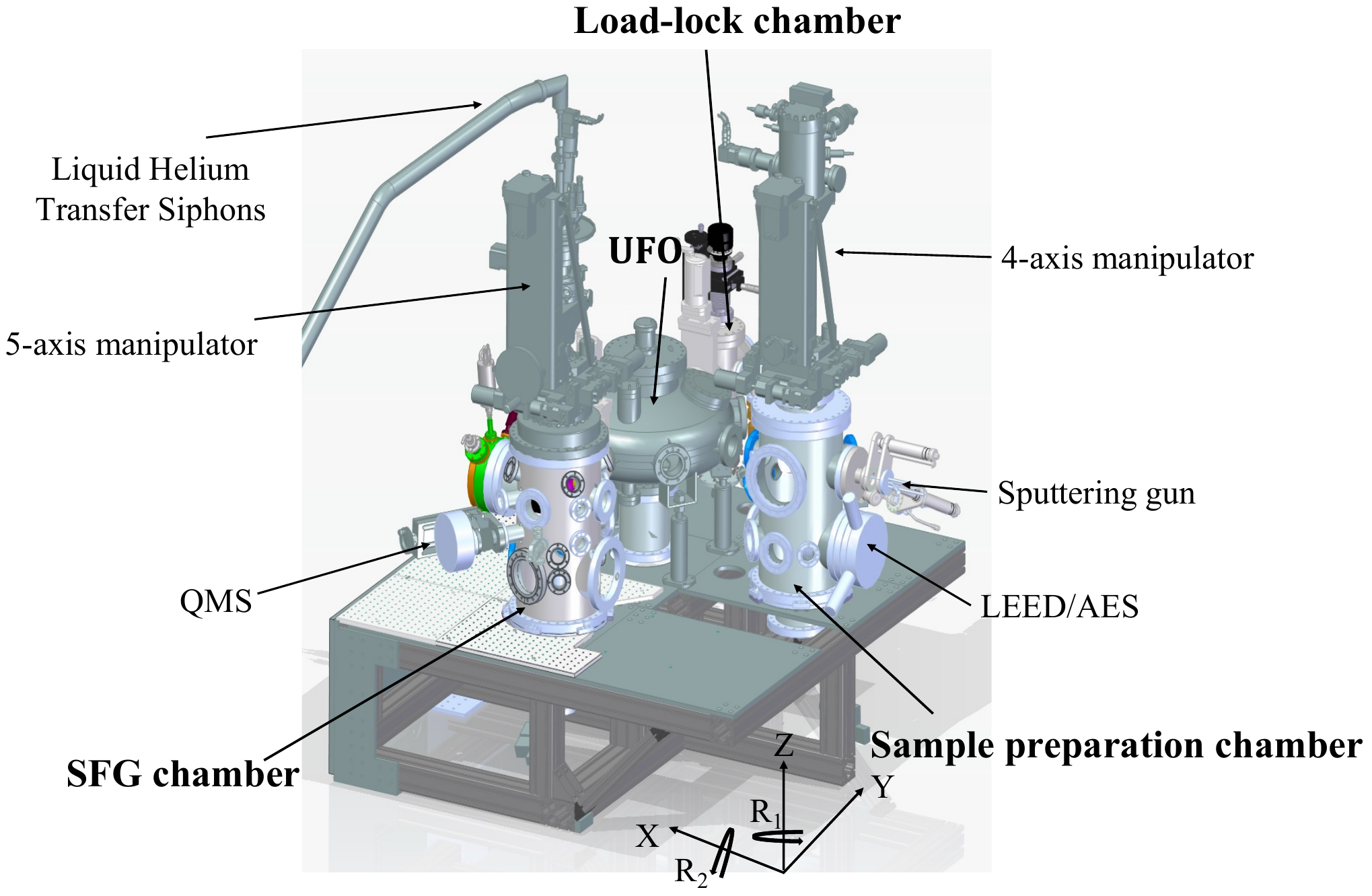}
	\caption{Schematic drawing of the ultra-high vacuum setup. Major components of the system are labeled. See text for details.}
	\label{fig:uhv}
\end{figure}

A schematic drawing of our ultra-high vacuum (UHV) setup is shown in~\autoref{fig:uhv}. 
It consists of four main chambers: a load-lock chamber for sample loading, a radial distribution (UFO) chamber (PREVAC GmbH) for sample transfer, a preparation chamber for sample preparation and inspection, and the SFG chamber for sample characterization. 
The setup allows in-vacuum sample preparation, transfer, and characterization. 
The load-lock chamber has an O-ring sealed quick access door and a stainless steel frame holder for loading and storing a sample. 
The sample preparation chamber contains an argon ion sputtering gun (SPECS GmbH), a Low Energy Electron Diffraction/Auger Electron Spectroscopy instrument (SPECS GmbH), a quadrupole mass spectrometer (RGA100, Stanford Research Systems) for Temperature Programmed Desorption Spectroscopy, and a four-axis manipulator (PREVAC GmbH) which can translate the sample along the x-, y- and z- axes and rotate it with respect to the z-. 
The manipulator can be cooled with liquid nitrogen to $\sim$\SI{-120}{\celsius} and heated to a temperature of $\sim$\SI{1100}{\celsius} by resistive or $\sim$\SI{1400}{\celsius} by electron beam heating. 
An argon ion sputtering gun is used for cleaning the sample surface by bombarding it with high-voltage (\textit{i.e.}\,1~kV) accelerated argon ions. 
The sample preparation chamber is mounted with a leak valve (VAT Group AG) that enables controlled dosing of a wide range of gases. 
The radial distribution chamber is located at the center of the setup, contains a rotary transfer arm and is connected with the other three chambers by gate valves (VAT Group AG); a configuration chosen so as to allow separate pumping of each chamber.
This configuration allows independent pumping of each chamber and transfer of samples between all chambers without leaving vacuum. 
The SFG chamber is equipped with a quadrupole mass spectrometer (RGA200, Stanford Research Systems) for Temperature Programmed Desorption Spectroscopy, a five-axis motorized manipulator (PREVAC GmbH) that can translate the sample along the x-, y- or z-axes and and rotate it with respect to the z-axis and the sample surface normal (x- or y-axis depending on sample orientation). 
The sample can be cooled to \SI{-220}{\celsius} with liquid helium or \SI{-120}{\celsius} with liquid nitrogen and heated to $\sim$\SI{1100}{\celsius} by resistive or $\sim$\SI{1400}{\celsius} by electron beam heating. 
The SFG chamber is mounted with two independent leak valves (VAT Group AG) that enable \textit{in-situ} SFG measurements while gas dosing with controlled dosing pressures. 
To achieve an ultra-high vacuum pressure, each chamber is evacuated by an independent oil-free turbo molecular pump (Pfeiffer Vacuum) backed by a HiCube 80 Eco turbo pumping station (Pfeiffer Vacuum). 
The typical pressure inside each chamber (excepting the load-lock) is $\sim 1\times10^{-10}$ mbar.

\subsection{Description of our laser setup}
\label{sec:laser}
\begin{figure*}[!htbp]
	\centering
	\includegraphics[width=0.68\textwidth]{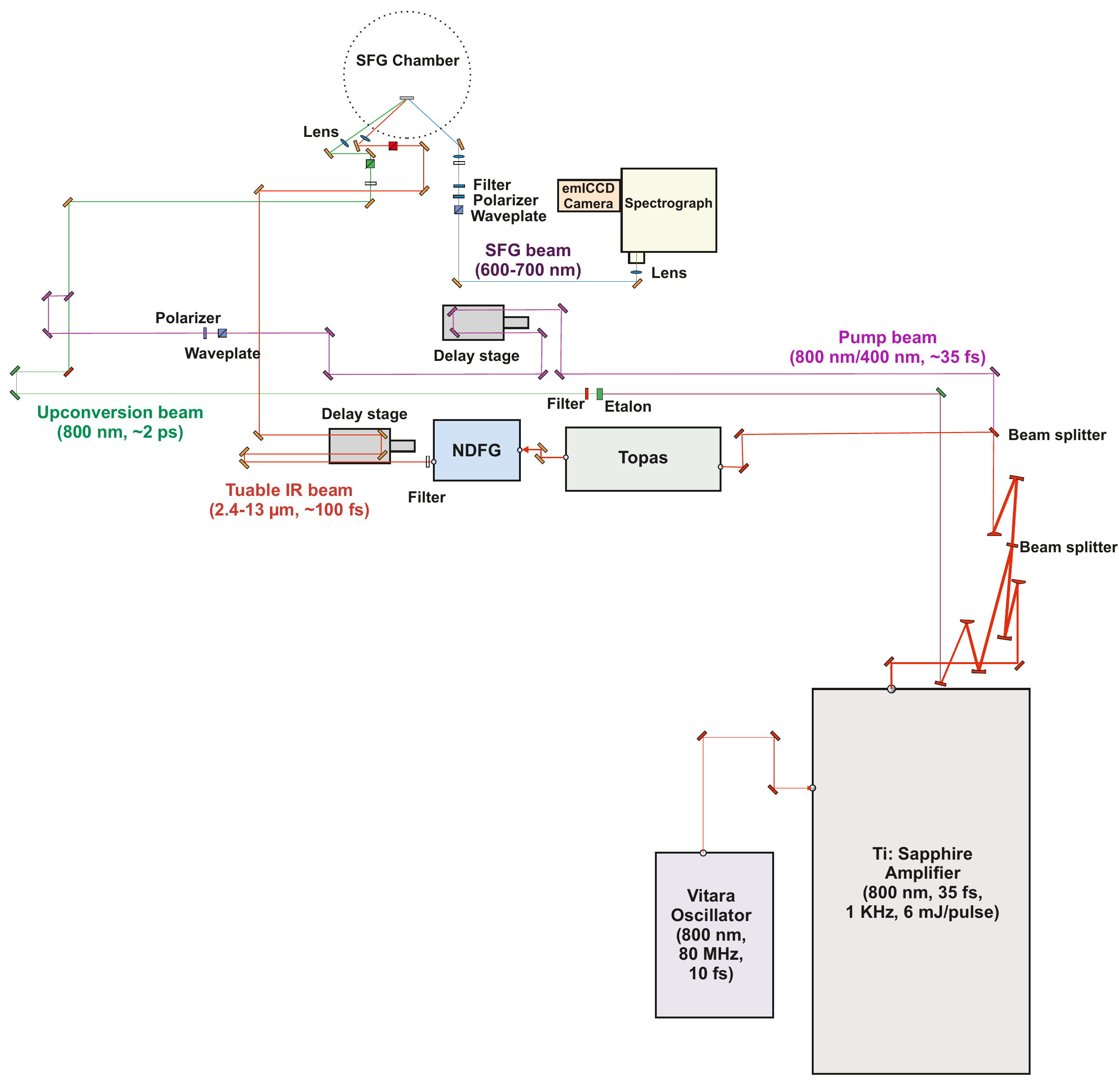}
	\caption{Schematic drawing of the femtosecond laser setup for (time-resolved) sum frequency generation spectroscopy measurement. See text for details.}
	\label{fig:laser}
\end{figure*}
A schematic drawing of the laser setup is shown in~\autoref{fig:laser}. 
We employ a laser system composed of a Ti: Sapphire oscillator (Vitara, Coherent) that seeds a regenerative amplifier (Legend Elite Duo HE + USP, Coherent) that delivers pulses with an energy of $6\frac{\text{mJ}}{\text{pulse}}$, a center wavelength of 800 nm, a duration of 35~fs (FWHM) and a repetition rate of 1 KHz. 
The amplifier output is split into three beams. 
One beam with a power of $\sim$1.8~W (1.8~mJ/pulse) pumps a commercial optical parametric amplifier (TOPAS-Prime, Light Conversion). 
The signal and idler beams produced by the OPA are subsequently mixed in a collinear difference frequency generation scheme to generate broadband infrared (IR) femtosecond pulses, the center wavelength of which can be tuned from 2-13~\textmu m. 
The second beam is spectrally shaped by an air-spaced etalon (SLS Optics Ltd) to produce a narrowband 800 nm beam.
To conduct vibrationally or final-state resonant SFG measurements the infrared and narrow-band 800 nm beams are spatially and temporally overlapped at the sample surface. 
The third beam propagating on this portion of the table is part of the 800 nm femtosecond amplifier used as a \emph{pump} in various pump - SFG probe experiments. 
The 800 nm pump is straightforwardly double or tripled in frequency using non-linear crystals (\textit{e.g.}\ BaB$_{2}$O$_{4}$, LiB$_{3}$O$_{5}$, KH$_{2}$PO$_{4}$, KTiOPO$_{4}$, etc.) to generate alternative UV pump pulses at 400 nm or 267 nm~\cite{Boyd:SHG:1968, Seka:THG:1980}.   
The SFG chamber is designed so as to allow the pump beam to propagate collinearly with the narrowband 800 nm beam as shown in ~\autoref{fig:laser} or to pump over a range of possible angles of 0$\sim$5\textdegree~(but within the plane defined by the incident narrow band 800 and infrared). 
This collinear geometry makes it straightforward to conduct pump-SFG probe measurements without the need to change the SFG beam path. 
The relative delay between the pump pulse and the IR or narrowband 800 nm pulse is controlled by a motorized optical delay stage (GTS150, Newport Corporation) in the optical path of the pump. 
A $\frac{\lambda}{2}$ waveplate and a polarizer are mounted in each of these three beams to allow continuous pulse energy adjustment while maintaining a clean p- or s- polarization. 
The diameter of the cylindrical SFG chamber is $\sim$ 30 cm, thus the focal distance of the lens for the incident three beams (pump, narrowband 800 and tunable IR) need to be longer than 15 cm (given a sample located at the center of the chamber). 
Here, the incident IR beam is focused by an uncoated ZnSe lens with a focal length of 25 cm. The narrowband 800 and pump beams are focused by a N-BK7 plano-convex lens with a focal length of 40 cm. 
In a configuration in which all three beams enter the SFG chamber through a single viewport, a DN63CF KBr window (Torr Scientific Ltd) is employed.
The incidence angles of the IR beam and the narrowband 800 nm beam with respect to the surface normal axis in this single viewport configuration are 46.1\textdegree~and 53.1\textdegree~respectively. 
The focus beam diameters (after projecting on the sample) of the narrowband 800 and infrared beams are $\sim$ 110~\um and 270~\um (for the long-axis), respectively. 
The dimensions of this spot can be readily decreased in the current set up. 
A lens with a focal length of 15 cm (the smallest distance compatible with the 15 cm radius of the SFG chamber) would decrease spot size (short axis) to 41~\um. 
Similarly, doubling the visible beam with a telescope before focussing would allow a further decrease in spot size to 21~\um.
Once the two beams are spatially and temporally overlapped on the sample surface, an SFG beam will be produced.
The generated SFG beam is collimated by a 40 cm lens and then filtered by a short pass filter with a cutting wavelength at 750 nm to eliminate the narrow-band 800 nm and pump pulses. 
The filtered SFG beam is propagated through a polarizer, to select only clean p- or s- polarized SFG emission, then rotated to p- polarization by a $\frac{\lambda}{2}$ waveplate and dispersed by a spectrograph (SpectraPro HRS-300) onto an emICCD camera (Princeton Instrument PI-MAX 4). 
Ensuring all SFG fields are p- polarized beam before entering the spectrograph removes the necessity to correct for grating efficiencies that are a function of field polarization ~\cite{Hui:grating:2023}. 
The polarization dependent SFG measurement is realized using different polarization combinations of the narrow-band 800 nm and IR beams and detecting the generated SFG of different polarization. 

Much prior work has shown that polarization resolved detection of vibrationally resonant SFG can be used to determine the orientation of adsorbed molecules on solid surfaces and interface-induced intermolecular coupling \cite{zwa18}. 
In the next section we demonstrate this capability in our set-up by examining the prototypical system CO on Pt(111) (see SI for laser parameters).

\subsection{Polarization Resolved VSFG of CO on Pt(111) in UHV}
\label{sec:PSFG}
\begin{figure*}[!htbp]
	\centering
	\includegraphics[width=0.65\textwidth]{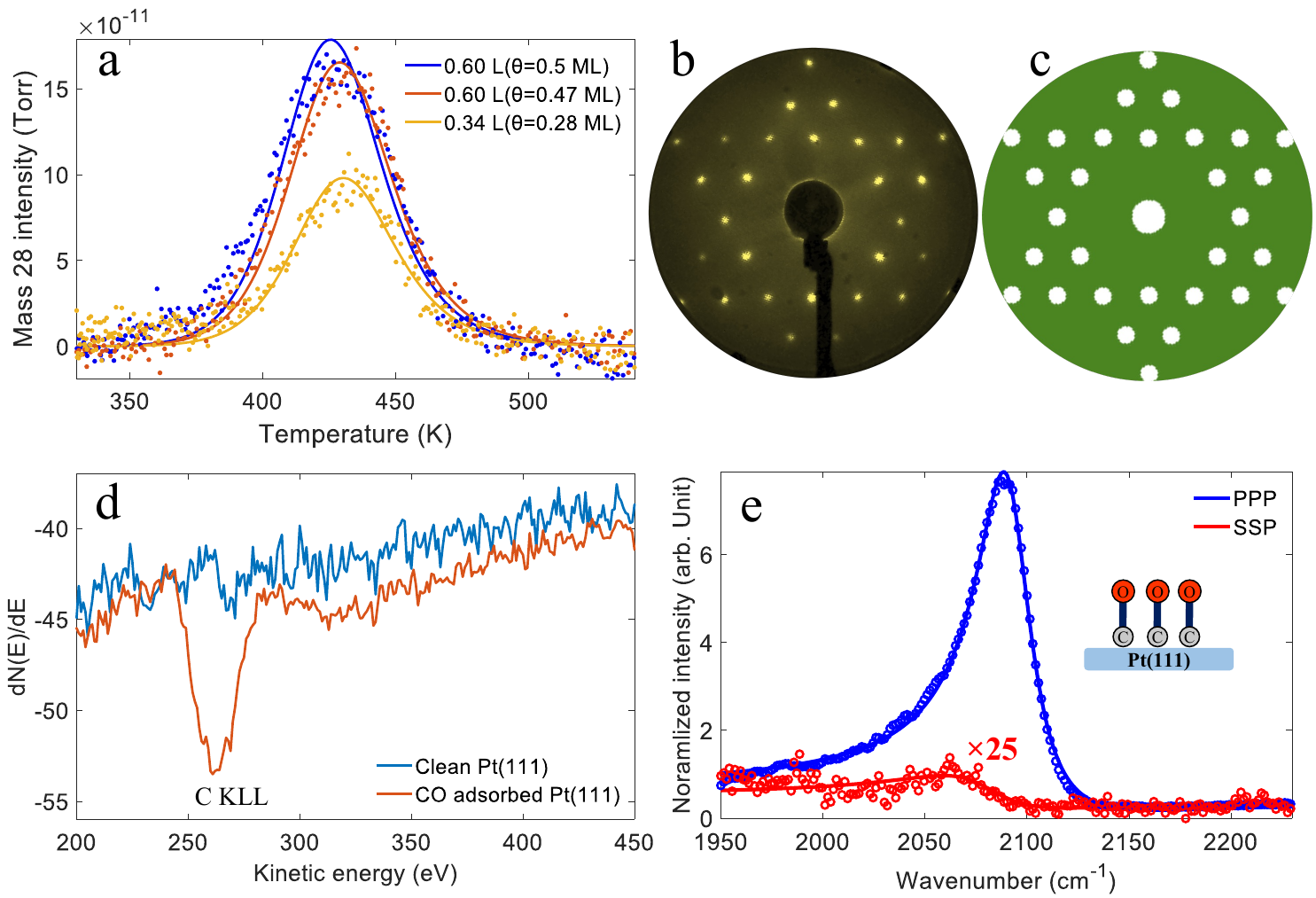}
	\caption{(a) Temperature programmed desorption spectrum of CO chemisorbed on Pt(111) with three different CO dosing quantities: 0.84 L (blue), 0.60 L (red), and 0.34 L (orange).
    (b) Low energy electron diffraction pattern of CO adsorbed on Pt(111) with a coverage of 0.5 monolayer. The probe electron beam energy is 32~eV. (c) Simulated LEED result of a c(4$\times$2) adsorbed structure on a Pt(111) surface (see SI for details). 
    (d) Auger electron spectrum of a clean Pt(111) surface (blue) and a CO chemisorbed Pt(111) surface (red). (e) Vibrationally resonant sum-frequency spectrum of chemisorbed CO on Pt(111) with a coverage of 0.5 ML under \textit{ppp} polarization (blue) and \textit{ssp} polarization (red). See text for details.}
	\label{fig:CO}
\end{figure*}

Platinum (Pt) is widely employed as a catalyst for (de)hydrogenation, reduction and oxidation reactions due to its high activity, stability, and selectivity. 
Reactions on Pt involving the transformation of small hydrocarbons often produce CO as an intermediate \cite{fre11,bae22}. 
As a result the adsorption of CO on Pt(111) has been studied extensively.~\cite{Ertl:CO:1977,Steininger:CO:1982} 
We prepared clean Pt(111) surfaces in the sample preparation chamber by Ar ion bombardment (with an acceleration voltage of 1000~V and filament current of 10~mA) for one hour followed by annealing at  \SI{900}{\celsius} for 10 minutes with the sputtering$\rightarrow$annealing cycle repeated five times.
The heating rate to reach the annealing temperature was $\approx 300 \frac{\text{\SI{}\celsius}}{\text{min}}$ and the cooling rate $\approx -250 \frac{\text{\SI{}\celsius}}{\text{min}}$.
After these sputtering and annealing cycles, we verified that the Pt(111) surface was contaminant free by Auger electron spectroscopy (see \autoref{fig:CO}(d)) and that the surface has the expected $1\times 1$ LEED pattern (see Supporting Information).
Having verified that the sample surface was contaminant free and had the expected structure, the sample was cooled to \SI{-70}{\celsius} for CO dosing. 
CO dosing was performed via a leak valve.
We calibrated CO surface coverage to dosing conditions by dosing for a defined time and pressure, shutting the leak valve and collecting temperature programmed desorption (TPD) spectra. 
TPD spectra were collected by heating the sample with a rate of 80 \SI{}\celsius/min and detecting the increase in CO partial pressure (as CO desorbs from the surface) with a quadrupole mass spectrometer whose tip is placed 3~mm away from the sample surface. 
Since the pumping speed of our UHV setup is much larger than the CO partial pressure increase due to the desorption, the measured CO partial pressure is proportional to the desorption rate \footnote{The pumping speed of the HiPace 700 H turbomolecular pump is 685 L/s for N$_{2}$}.
This measurement allows calibration of the TPD response of a CO saturated monolayer (multilayer adsorption is evident as a lower temperature feature in the TPD spectra that grows in under extended dosing).  

\autoref{fig:CO}(a) shows the TPD spectrum of CO on Pt(111) with three different CO surface coverages.
Following prior workers we assume desorption of CO from Pt(111) during a thermal ramp has a single rate limiting step and extract the desorption energy ($\text{E}_{\text{CO}}$) and order ($x$) from the data by fitting the results with the Polyani-Wigner equation ~\cite{Polanyi:TPD:1925,King:TPD:1975,Jong:TPD:1990}, 
\begin{equation}\label{e:Polayani}
	-\frac{d\Theta_{\text{CO}}}{dt} = \nu_{\text{CO}}\Theta_{\text{CO}}^{x}\exp\left(-\frac{\text{E}_{\text{CO}}}{\text{RT}}\right)
\end{equation}	
where $\Theta_{\text{CO}}$ is the surface coverage of CO and $\nu_{\text{CO}}$ is the preexponential frequency factor.
We find $\text{E}_{\text{CO}}$ on Pt(111) to be 110~kJ/mol and the desorption to have an order of 2 consistent with previous results~\cite{Ertl:CO:1977, Steininger:CO:1982}. 
Since it is known that the saturation coverage of CO on Pt(111) in UHV at room temperature is 0.5 ML~\cite{Norton:Coscoverage:1982, Ertl:CO:1977},
the integrated area of the TPD spectrum at saturated CO monolayer coverage allows quantification of surface coverage under reduced dosing. 
\autoref{fig:CO} (b) shows the LEED pattern of CO adsorbed on Pt(111) with a coverage of 0.5 ML with the probe electron beam energy of 32~eV. 
\autoref{fig:CO} (c) shows the LEEDpat~\cite{Hermann:LEED:2022} simulated result from a c(4$\times$2) adsorption structure.
The calculated structure is clearly consistent with the experimental results. 
Viewed in real-space, an 0.5 ML coverage and c(4$\times$2) adsorbate structure is consistent with a structure in which adsorbed CO molecules populate both on-top and bridge sites
of Pt(111) ~\cite{Steininger:CO:1982}. 
\autoref{fig:CO} (d) shows the Auger electron spectra of the clean Pt(111) surface (blue) and the same surface with a half monolayer of adsorbed CO (orange). 
Clearly the C KLL peak~\cite{Maglietta:AES:1975}, centered at around 262~eV, appears only after CO dosing.

Collecting vibrationally resonant sum frequency spectra requires spatially and temporally overlapping incident infrared and visible fields at a sample surface and detecting the intensity of the emitted field at the sum of the two incident frequencies.
\autoref{fig:CO}(e) shows such a VSFG spectrum (a plot of measured $I_{\text{sf}}$ vs.\ $\omega_{\text{ir}}$ over 1950-2200 cm$^{\textit{-1}}$) collected from a sample with a half-monolayer ($\Theta$ = 0.5 ML) CO adsorbed on Pt(111) collected under the \textit{ppp} (blue) and \textit{ssp} polarization conditions (red) measured in the SFG chamber (where \textit{ssp} indicates \text{s}-polarized SFG, \textit{s}-polarized narrowband 800 nm and \textit{p}-polarized IR.
\textit{s} indicates $\perp$ to the incident plane and \textit{p} $\parallel$).  
On-top adsorbed CO on a wide variety of metals is known to be resonant at the infrared photon energies we employ \cite{bra70,tre00,zae12}.

The ratio of vibrationally resonant SFG intensity measured under the \textit{ppp} to that measured under the \textit{ssp} condition, \textit{i.e.}\ $\frac{I_{\text{sf,ppp}}}{I_{\text{sf,ssp}}}$, is a function of the orientation of CO on the Pt(111) surface and the ratio of components of the hyperpolarizability tensor, \textit{i.e.}\ the second order molecular response $\beta^{(2)}$ \cite{Lambert:SFG:2005,Zhuang:PSFG:1999,Li:CO:2018}. 
While $\beta^{(2)}$ is a function of local environment, particularly near resonant optical transitions, its value is rarely accessible in condensed molecular phases or at interfaces making it challenging to separate the contribution of orientation and electronic structure in the observed sum frequency response~\cite{Hirose:SFGtheory:1992, Zhuang:PSFG:1999, Wang:SFG:2005, Lambert:SFG:2005}.
As noted above, prior work has demonstrated that the sharply resolved c(4$\times$2) LEED pattern of CO adsorbed on Pt(111) strongly suggests that atop adsorbed CO is oriented normal to the Pt(111) surface~\cite{Apai:orientCOarps:1976, Hofmann:orientCOarpes:1982, Wesner:orientCOxps:1986, Ogletree:orientCOleed:1986}.
Thus this system offers a relatively rare opportunity to elucidate the $\beta^{(2)}$ for CO adsorbed on Pt(111) at a coverage of 0.5 ML. 
To understand how the macroscopic observables are related to microscopic structure requires review of the physical basis of the VSFG response \cite{Lambert:SFG:2005,Zhuang:PSFG:1999}.

The VSFG spectrum of CO adsorbed on Pt(111) shown in \autoref{fig:CO}(e) results from spatially and temporally overlapping a narrowband pulse train in the visible and an infrared pulse train in the mid-infrared,  
\begin{equation}
	\label{eq:SFG}
	I_{\text{sf}}(\omega_{\text{sfg}}) \propto \left|\chi^{(2)}_\text{eff}\right|^2 I_{\text{vis}}(\omega_{\text{vis}})I_{\text{ir}}(\omega_{ir})
\end{equation}
where $I_{\text{sf}}$ is the intensity of the emitted SFG field (and is a function of photon energy, \textit{i.e.}\ $\omega_{\text{SFG}}$), $I_{vis}$ is the intensity of the narrow band visible up-conversion pulse (and is a function of $\omega_{\text{vis}}$) and $I_{\text{ir}}$ is the intensity of the incident infrared field (and is a function of $\omega_{\text{ir}}$). 
$\chi^{(2)}_{\text{eff}}$, the \textit{effective} second order macroscopic susceptibility, is a 
the sum of nonresonant and vibrationally resonant transitions.
If each vibrational transition is only homogeneously broadened, it can be written,
\begin{equation}\label{eq:chi2}
	\chi^{(2)}_{\text{eff}}(\omega_{\text{ir}}) = \chi^{(2)}_\text{nr} + \chi^{(2)}_\text{res} = A_\text{nr}e^{i\phi_{\text{NR}}} + \sum_{q} \frac{A_{q}}{\omega_{\text{ir}} - \omega_q + i \Gamma_q}
\end{equation}
where $\chi^{(2)}_\text{nr}$ is the nonresonant (often assigned to the distant tail of high energy optically accessible transitions \cite{bus09}) and $\chi^{(2)}_\text{res}$ is the resonant part of the second-order nonlinear susceptibility.
$A_\text{nr}$ is the amplitude of the nonresonant background and $\phi_{\text{NR}}$ the phase.  
${A_{q}}$, $\omega_q$, $\Gamma_q$ are the amplitude, resonant frequency, and damping coefficient of the  $q^{\text{th}}$vibrational mode, respectively. 

$\chi^{(2)}_{\text{eff}}$ is \textit{effective} in \autoref{eq:SFG} in that it is also a function of the polarizations or angles (with respect to the surface normal) of the incident and emitted fields. 
Disentangling these parameters gives,
\begin{widetext}
\begin{eqnarray}\label{eq:SFGP}
 	\chi^{(2)}_\text{eff,ppp} &  = & -L_{\mathrm{XX}}(\omega_\text{sfg})L_{\mathrm{XX}}(\omega_\text{vis})L_{\mathrm{ZZ}}(\omega_\text{ir}) \cos\alpha_{\text{sfg}}\cos\alpha_{\text{vis}}\text{sin}\alpha_{\text{ir}}\chi^{(2)}_{\mathrm{XXZ}} \nonumber\\ 
	& & -L_{\mathrm{XX}}(\omega_\text{sfg})L_{\mathrm{ZZ}}(\omega_\text{vis})L_{\mathrm{XX}}(\omega_\text{ir})\cos\alpha_\text{sfg}\sin\alpha_{\text{vis}}\cos\alpha_{\text{ir}}\chi^{(2)}_{\mathrm{XZX}} \\
	& & +L_{\mathrm{ZZ}}(\omega_\text{sfg})L_{\mathrm{XX}}(\omega_\text{vis})L_{\mathrm{XX}}(\omega_\text{ir})\sin\alpha_{\text{sfg}}\cos\alpha_{\text{vis}}\cos\alpha_{\text{ir}}\chi^{(2)}_{\mathrm{ZXX}} \nonumber\\ 
	& & +L_{\mathrm{ZZ}}(\omega_\text{sfg})L_{\mathrm{ZZ}}(\omega_\text{vis})L_{\mathrm{ZZ}}(\omega_\text{ir})\sin\alpha_{\text{sfg}}\sin\alpha_{\text{vis}}\sin\alpha_{\text{ir}}\chi^{(2)}_{\mathrm{ZZZ}} \nonumber \\
 	\chi^{(2)}_\text{eff,ssp}  & = & L_{\mathrm{YY}}(\omega_\text{sfg})L_{\mathrm{YY}}(\omega_{\text{vis}})L_{\mathrm{ZZ}}(\omega_{\text{ir}})\sin\alpha_{\text{ir}}\chi^{(2)}_{\mathrm{YYZ}} \label{eq:SFGS}
\end{eqnarray}
\end{widetext}
where $\alpha_{i}$ denotes the angle of the \textit{i}$^{\text{th}}$ beam with respect to the surface normal. 
$\chi^{(2)}_{\mathrm{IJK}}$ is the macroscopic second-order susceptibility in laboratory coordinates in which $Z$ is the surface normal and all beams propagate in the $X-Z$ plane. 
$L_{II}(\omega_{i})$ denotes the Fresnel factor at frequency $(\omega_\text{i})$ and corrects for linear optical effects on the interfacial nonlinear optical response. 
The Fresnel factors can be written \cite{Lambert:SFG:2005,Zhuang:PSFG:1999},
\begin{eqnarray}\label{eq:fresnel}
	L_{\mathrm{XX}}(\omega) & = & \frac{2n_{\text{vac}}(\omega)\text{cos}~\gamma}{n_{\text{vac}}(\omega)\text{cos}~\gamma+n_{\text{Pt}}(\omega)\text{cos}~\alpha} \nonumber\\
	L_{\mathrm{YY}}(\omega) & = & \frac{2n_{\text{vac}}(\omega)\cos\alpha}{n_{\text{vac}}(\omega)\cos\alpha + n_{\text{Pt}}(\omega)\cos\gamma} \\
	L_{\mathrm{ZZ}}(\omega) & = & \frac{2n_{\text{vac}}(\omega)\cos\alpha}{n_{\text{vac}}(\omega)\cos\gamma+n_{\text{Pt}}(\omega)\cos\alpha}{\left(\frac{n_{\text{vac}}(\omega)}{n^{'}(\omega)}\right)}^2 \nonumber
\end{eqnarray}
where $n_{\text{vac}}$, $n_{\text{Pt}}$ and $n^{'}$ are the refractive indices of the vacuum, single crystal, and interfacial layer, respectively. $\gamma$ is the refracted angle $[n_{\text{vac}}(\omega)\sin\alpha = n_{\text{Pt}}(\omega)\sin\gamma]$. 
We estimated $n^{'}$ using a modified Lorentz model~\cite{Zhuang:PSFG:1999}, $n^{'} = n_{\text{vac}}n_{\text{Pt}}\sqrt{\frac{6+(n_{\text{Pt}})^{2}-(n_{\text{vac}})^{2}}{4(n_{\text{Pt}})^{2}+2(n_{\text{vac}})^{2}}}$. 
The refractive indices of the vacuum, single crystal Pt(111), and interfacial layer under the wavelengths of our experimental conditions are listed in the supplementary information (Table S1).

The macroscopic second-order susceptibility $\chi^{(2)}_{IJK}$ in laboratory coordinates is related to the microscopic hyperpolarizability tensor $\beta^{(2)}_{ijk}$ in the molecular coordinate system,
\begin{equation}\label{e:hyperpol}
	\chi^{(2)}_{\text{IJK}} = N_{\mathrm{s}}\sum_{\mathrm{i,j,k}} \left\langle \left(\hat{I}\cdot\hat{i}\right)\left(\hat{J}\cdot\hat{j}\right) \left(\hat{K}\cdot\hat{k} \right)   \right\rangle \beta^{(2)}_{\mathrm{ijk}}
\end{equation}
where $N_{\mathrm{s}}$ is the density of molecules in the focal spot and $\langle ... \rangle$ indicates an ensemble average of the Euler matrix necessary to transform individual molecules into the laboratory reference frame.
For CO, a molecule with C$_{\infty\text{v}}$ symmetry around the CO bond, $\beta^{(2)}_{aac} = \beta^{(2)}_{bbc}$.
Given this symmetry $\chi^{(2)}_{\mathrm{IJK}}$ is related to $\beta^{(2)}_{\mathrm{ijk}}$ \cite{Lambert:SFG:2005,Zhuang:PSFG:1999},
\begin{widetext}
\begin{eqnarray}\label{e:hyper_v_macro}
	\chi^{(2)}_{XXZ} = \chi^{(2)}_{YYZ} & = & \frac{1}{2}N_{\mathrm{s}}\beta^{(2)}_{ccc}\left[(1+R)\langle\cos\theta \rangle -(1-R)\langle\cos\theta\rangle^3 \right] \nonumber\\
	\chi^{(2)}_{XZX} = \chi^{(2)}_{ZXX} & = &\frac{1}{2}N_{\mathrm{s}}\beta^{(2)}_{ccc}\left[(1-R)\langle\cos\theta\rangle -(1-R)\langle\cos\theta\rangle^3\right]\\
	\chi^{(2)}_{ZZZ} & = & N_{\mathrm{s}}\beta^{(2)}_{ccc}\left[R\langle\cos\theta\rangle + (1-R)\langle\cos\theta\rangle^3\right] \nonumber
\end{eqnarray}
\end{widetext}
where $\theta$ is the orientation angle between the CO molecular axis and the surface normal, and R is the hyperpolarizability ratio ($R = \beta^{(2)}_{aac}/\beta^{(2)}_{ccc}$).
A related quantity, the bond polarizability ratio ($r$) is often relevant in Raman spectroscopy. 
For isotropic systems the measurable Raman depolarization ratio ($\rho$) is related to the bond polarizability ratio ($r$) $\rho = \frac{3}{4+5[(1+2r)/(1-r)]^2}$~\cite{Wang:SFG:2005}. 
For molecules, such as CO, with C$_{\infty \text{v}}$ symmetry the hyperpolarizability ratio and the bond polarizability ratio are equivalent.

Given an $R$ measured in bulk solution, \autoref{e:hyper_v_macro} allows, provided the vibrational response of CO is independent of local environment, direct determination of interfacial CO orientation from measured VSFG spectra. 
This independence is, however, not generally true for interfaces of catalytic interest (\textit{e.g.}\ CO shows a large electrochemical Stark shift \cite{lam96}). 

Previous workers have found an $R$-value of the free CO molecule from density functional theory of 0.25 \cite{Gisbergen:PSFG:1998}. 
In contrast, experimentally, the $R$-value of on-top adsorbed CO on Pt(111) has been found to be 0.6 at an electrode aqueous electrolyte interface \cite{Baldelli:PSFG:1999} and 0.49 for multilayer CO at Pt(111) gas phase interface\cite{Li:CO:2018}. 
Both efforts calculated $R$ from polarization dependent SFG measurements assuming CO molecules oriented normal to the surface.
As noted above, the well-defined c(4$\times$2) pattern we observe in LEED is strong independent evidence that CO is oriented normal to the surface for (sub)monolayer coverages of CO on Pt(111) over a wide range of temperature ~\cite{Ogletree:orientCOleed:1986}.
This system thus offers an opportunity to determine the $R$ value of interfacial CO without assumptions about orientation. 

As is perhaps obvious from equations \ref{eq:chi2}, \ref{eq:SFGP} and \ref{eq:SFGS}, extracting either $R$ or $\theta$ from arbitrary VSFG spectra measured under the \textit{ppp} and \textit{ssp} polarization conditions is not generally possible.
Taking the \emph{ratio} of $I_{\mathrm{sf,ppp}}$ to $I_{\mathrm{sf,ssp}}$ helps, $N_{\mathrm{s}}$ cancels, but leaves the problem of correctly describing the four different possible contributions to $\chi^{(2)}_{\mathrm{eff,ppp}}$. 
Because our measured spectrum have only a single resonance, and $\chi^{(2)}_{\text{res}} \gg \chi^{(2)}_{\text{nr}}$, substantial simplifications are possible.
On the LHS of \autoref{eq:SFGP} $\chi^{(2)}_{\mathrm{eff,ppp}} \approx \chi^{(2)}_{\mathrm{res,ppp}}$ and both the LHS and the RHS can be evaluated at the single, maximum, measured intensity with no introduction of error. 
Pursuing such an approach, and making a similar approximation in analyzing the $\textit{ssp}$ spectra, the measured $\frac{I_{\mathrm{sf,ppp}}}{I_{\mathrm{sf,ssp}}} = 182.6$ implying $R=0.07$.
The orientation of on-top adsorbed CO on Pd(100) is known to deviate from normally oriented with multilayer formation due to adsorbate/adsorbate interaction \cite{Ouvrard:CO:2014}.
The lower $R$ we observe relative to that Li et al \cite{Li:CO:2018} infereed under high CO coverage, is thus consistent with a scenario in which $R$ is CO coverage dependent or orientation at higher pressures deviates from normal. 
Similarly the lower $R$ we observe, relative to that inferred by Baldelli et al.\ \cite{Baldelli:PSFG:1999}, assuming CO is normally oriented at a Pt(111) working electrode, is consistent with scenario in which $R$ depends on local environment, \textit{i.e.}\ in this case electrolyte or bias, or CO orientation is non-normally oriented at the electrochemical interface. 
Resolving this discrepancy for both environments is important in quantitatively inferring interfacial structure from nonlinear optical observables and is an object of current research~\cite{TONG2024}.

\subsection{Temporal Resolution of the SFG Spectral Response}
\label{sec:TSFG}
\begin{figure}[!htbp]
	\centering
	\includegraphics[width=0.4\textwidth]{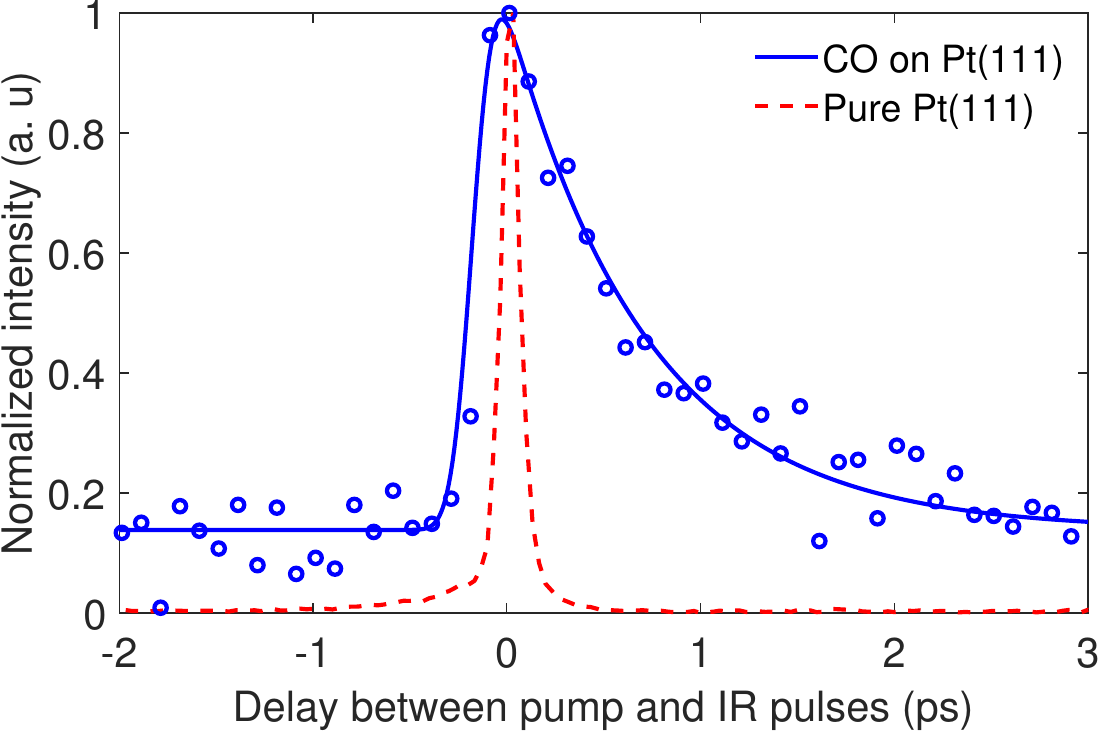}
	\caption{The red dashed line shows the cross-correlation of the femtosecond pump beam and IR beam. 
	The data is derived from integrating the measured sum frequency spectrum from clean Pt(111) as a function of the delay between the pump and IR pulses. 
	The blue curve shows the vibrational dephasing dynamics of CO after IR excitation. 
	The data is derived from integrating the measured sum frequency spectrum from 0.5~ML CO covered Pt(111) as a function of the delay between the two beams. 
	See text for details.}
	\label{fig:pumpsfg}
\end{figure}
Having confirmed that our set-up offers the capability of measuring time-averaged VSFG spectra of molecular adsorbates on single crystals~\cite{Klunker:CO:1996, Roeterdink:COSFG:2005, Li:CO:2018})  
we next show, as initially demonstrated by previous workers~\cite{Bonn:TSFG:2000, Fournier:UHVSFG:2004, bac05, Backus:UHVSFG:2007, Nagao:UHVSFG:2009, Arnolds:UHVSFG:2010}, 
that it also allows the resolution of energy flow dynamics on femtosecond timescales. 

\autoref{fig:pumpsfg} shows the spectrally integrated SFG signal derived from temporally and spatially overlapping a 35 fs femtosecond 800 nm pulse (the pump beam line discussed in \autoref{sec:laser}) and the IR probe beam with a center wavelength at 4725~nm as a function of the delay between the two pulses (dashed red line). 
Because electron relaxation in Pt is sub-10 fs and frequency independent at these photon energies, this signal is solely a function of the length and spectral shape of these two beams~\cite{Wolf:metalsurface:1997}.
The FWHM of the resulting cross-correlation is 80~fs: our temporal resolution, assuming both IR and 800 nm pulses are Gaussians in the time domain, is 34~fs ($\sigma \approx \frac{FWHM}{2.35}$).

The blue points in \autoref{fig:pumpsfg} show a similar signal now collected from a Pt(111) surface prepared as indicated above with 0.5 monolayer of CO.
Fitting the data with a single exponential (blue line in \autoref{fig:pumpsfg}) quantifies the vibrational dephasing time: 1.1~ps.
This result is both consistent with previous work ~\cite{Beckerle:CO:1991} and, after Fourier transformation, yields a frequency domain bandwidth of 15 cm$^{-1}$.
Our direct measurement of the line width in the frequency domain, shown in \autoref{fig:CO}, yields a linewidth of 16 cm$^{-1}$ after fitting with \autoref{eq:chi2}. 
While in principle sampling in the time- or frequency domains offers no additional insight, in practice the each scheme is often sensitive to different features of the spectral response \cite{rok03}. 
The developed setup provides options to detect both of them with high precision. 

As discussed in \autoref{sec:laser} the ability to resolve VSFG signals \emph{both} from combining the temporally short `pump' 800 nm and probe IR \emph{and} the temporally long `probe' 800 nm and probe IR suggests pump-probe time-resolved sum-frequency generation spectroscopy, in which either the dynamics of vibrational relaxation~\cite{Backus:UHVSFG:2007} or that of charge transfer between molecular adsorbates and the solid, initiated by the the femtosecond pump pulse should be straightforward. 
Such experiments are required to understand the elementary processes that underlie, catalysis, solar cells and a variety of sensing applications \cite{pon17,fre22}

\subsection{The Ag(111) azimuthal dependent SFG Response}
Having shown that time-averaged, polarization resolved VSFG spectroscopy and femtosecond time-resolved VSFG spectroscopy are possible within our set-up we next demonstrate the possibility of measuring the polarization resolved VSFG signal as a function of azimuthal angle. 

\begin{figure}[!htbp]
	\centering
	\includegraphics[width=0.5\textwidth]{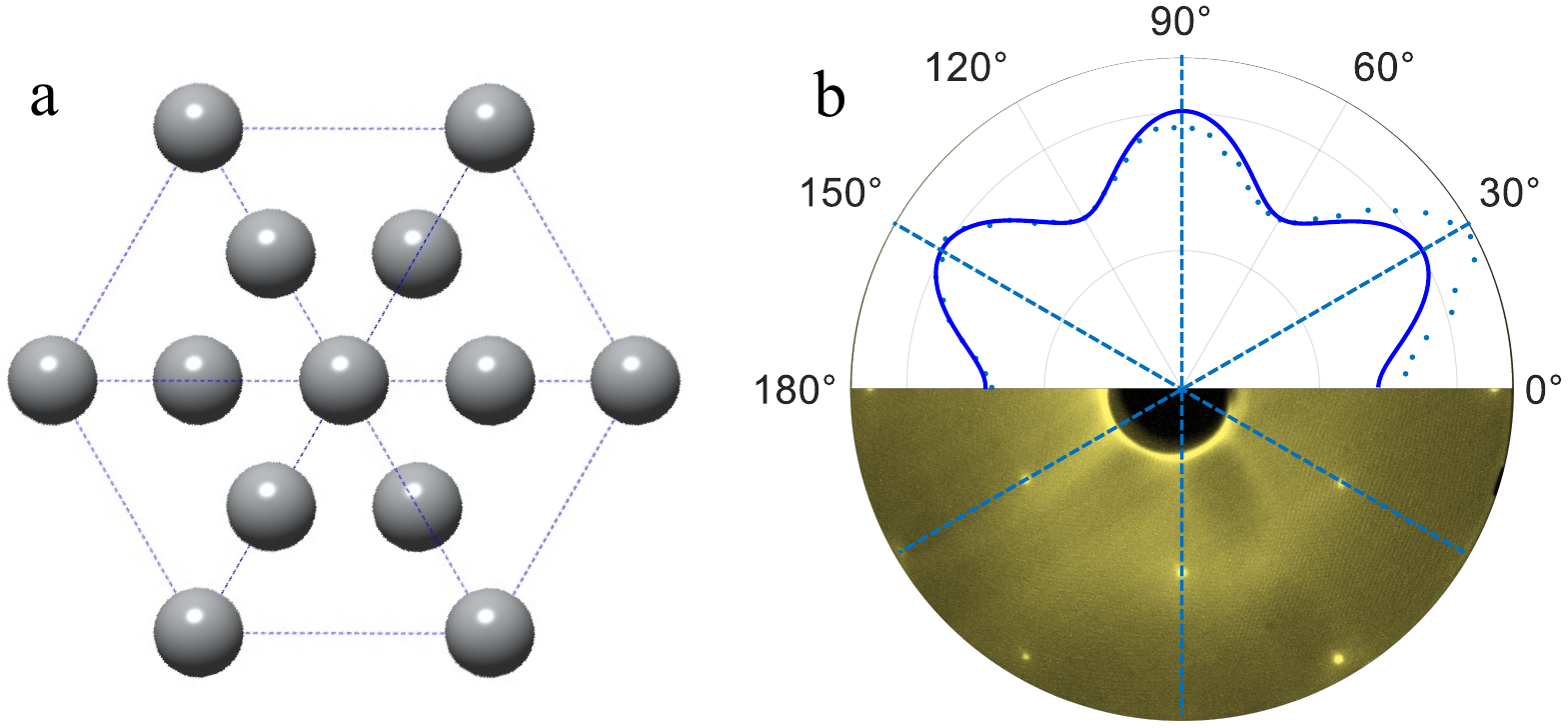}
	\caption{(a) The Ag(111) unit cell structure. (b) Top part: Integrated sum-frequency generation spectrum of Ag(111) as a function of azimuthal angle from 0$^{\circ}$ to 180$^{\circ}$. Bottom part: Low Energy Electron Diffraction (LEED) pattern of the Ag(111) surface collected with an electron beam energy of 145 eV. See text for details.}
	\label{fig:azimuth}
\end{figure}

Silver (Ag) is commonly used as a catalyst in producing ethylene oxide and formaldehyde~\cite{Wen:Ag:2014,Li:Ag:2019}.
Solid elemental Ag has a face-centered cubic lattice~\cite{Suh:Ag:1988}.
Thus its (111) surface exhibits a hexagonal close-packed structure (as shown in~\autoref{fig:azimuth}(a)), belongs to the C$_{3\text{v}}$ point group and, along the surface normal,  symmetry~\cite{Friedrich:AgASFG:1993} and, along the surface normal, has an ABCABC stacking sequence. 
Characterization of the interfacial nonlinear optical response vs.\ azimuthal angle offers the perspective of probing adsorbate orientation relative to Ag surface structure (as well as the opportunity discussed earlier to probe the SFG spectral response of TMDC monolayers at interfaces).

As discussed in \autoref{sec:UHV}, the SFG chamber is equipped with a high-stability motorized 5-axis UHV manipulator that, in principle, should enable SFG measurements that rely on rotating or translating the sample and also require in-situ heating or cooling the sample. 
To conduct the azimuthal resolved measurement we first prepare a clean Ag(111) sample using a sputtering annealing treatment in which sputtering is accomplished by Ar ion bombardment (with a filament current of 10~mA and acceleration voltage of 1000~V) for one hour followed by annealing at \SI{600}{\celsius} for 10 minutes~\cite{Engelhardt:Agcleaning:1976, Chelvayohan:Agcleaning:1982}. 
We found 5 such cycles in the sample preparation chamber to be sufficient to produce a clean, well-defined Ag(111) surface. 
Solid surfaces with C$_{3\text{v}}$ symmetry have 11 nonvanishing and 5 independent hyperpolarizability tensor elements.
For Ag(111) these are \textit{i.e.}\ $\beta_{bbb}^{(2)}=-\beta_{baa}^{(2)} = -\beta_{aab}^{(2)} =-\beta_{aba}^{(2)}$, $\beta_{aca}^{(2)} = \beta_{bcb}^{(2)}$, $\beta_{aac}^{(2)}=\beta_{bbc}^{(2)}$, $\beta_{caa}^{(2)}=\beta_{cbb}^{(2)}$,  $\beta_{ccc}^{(2)}$ where $a$, $b$ and $c$ are the crystallographic axes \cite{Boyd:ASFG:2020,Yang:ASFG:2023}.
The effective second-order nonlinear susceptibility measured under the \textit{ppp} polarization condition, \textit{i.e.}\ the $\chi^{(2)}_\text{ppp}$, is given by the following (where the $c$ crystallographic axis is parallel to the z-axis surface normal, both incident laser beams propagate in the x-z plane and the $a-b$ crystallographic plane is parallel to the x-y plane of the laboratory frame.). 
\begin{widetext}
\begin{eqnarray}\label{eq:SFGPc3v}
	\chi^{(2)}_\text{eff,ppp} & = & -L_\mathrm{xx}(\omega_\text{SFG})L_\mathrm{xx}(\omega_\text{Vis})L_\mathrm{xx}(\omega_\text{IR})\cos\alpha_\text{SFG}\cos\alpha_\text{Vis}\text{cos}~\alpha_\text{IR}\chi^{(2)}_\mathrm{xxx} \nonumber\\
    & & -L_{\mathrm{xx}}(\omega_\text{sfg})L_{\mathrm{xx}}(\omega_\text{vis})L_{\mathrm{zz}}(\omega_\text{ir}) \cos\alpha_{\text{sfg}}\cos\alpha_{\text{vis}}\text{sin}\alpha_{\text{ir}}\chi^{(2)}_{\mathrm{xxz}} \nonumber\\ 
	& & -L_{\mathrm{xx}}(\omega_\text{sfg})L_{\mathrm{zz}}(\omega_\text{vis})L_{\mathrm{xx}}(\omega_\text{ir})\cos\alpha_\text{sfg}\sin\alpha_{\text{vis}}\cos\alpha_{\text{ir}}\chi^{(2)}_{\mathrm{xzx}} \nonumber\\
	& & +L_{\mathrm{zz}}(\omega_\text{sfg})L_{\mathrm{xx}}(\omega_\text{vis})L_{\mathrm{xx}}(\omega_\text{ir})\sin\alpha_{\text{sfg}}\cos\alpha_{\text{vis}}\cos\alpha_{\text{ir}}\chi^{(2)}_{\mathrm{zxx}} \nonumber\\ 
	& & +L_{\mathrm{zz}}(\omega_\text{sfg})L_{\mathrm{zz}}(\omega_\text{vis})L_{\mathrm{zz}}(\omega_\text{ir})\sin\alpha_{\text{sfg}}\sin\alpha_{\text{vis}}\sin\alpha_{\text{ir}}\chi^{(2)}_{\mathrm{zzz}} \nonumber \\
    & = & -L_\mathrm{xx}(\omega_\text{SFG})L_\mathrm{xx}(\omega_\text{Vis})L_\mathrm{xx}(\omega_\text{IR})\cos\alpha_\text{SFG}\cos\alpha_\text{Vis}\cos\alpha_\text{IR}\beta_{bbb}\sin3\varphi \nonumber \\
     & & -L_{\mathrm{xx}}(\omega_\text{sfg})L_{\mathrm{xx}}(\omega_\text{vis})L_{\mathrm{zz}}(\omega_\text{ir}) \cos\alpha_{\text{sfg}}\cos\alpha_{\text{vis}}\text{sin}\alpha_{\text{ir}}\beta_{\mathrm{aac}} \nonumber\\ 
	& & -L_{\mathrm{xx}}(\omega_\text{sfg})L_{\mathrm{zz}}(\omega_\text{vis})L_{\mathrm{xx}}(\omega_\text{ir})\cos\alpha_\text{sfg}\sin\alpha_{\text{vis}}\cos\alpha_{\text{ir}}\beta_{\mathrm{aca}} \\
	& & +L_{\mathrm{zz}}(\omega_\text{sfg})L_{\mathrm{xx}}(\omega_\text{vis})L_{\mathrm{xx}}(\omega_\text{ir})\sin\alpha_{\text{sfg}}\cos\alpha_{\text{vis}}\cos\alpha_{\text{ir}}\beta_{\mathrm{caa}} \nonumber\\ 
	& & +L_{\mathrm{zz}}(\omega_\text{sfg})L_{\mathrm{zz}}(\omega_\text{vis})L_{\mathrm{zz}}(\omega_\text{ir})\sin\alpha_{\text{sfg}}\sin\alpha_{\text{vis}}\sin\alpha_{\text{ir}}\beta_{\mathrm{ccc}} \nonumber
\end{eqnarray}
\end{widetext}
in which $\varphi$ is the azimuthal rotational angle: the angle of the x-z plane with respect to the a-c around the z axis.
The measured $I_{\mathsf{sf}}$ under ppp polarization we thus expect to simplify to (where A and B are constants that are independent of orientation),
\begin{equation}
	\label{eq:ASFG}
	I_{\text{sf,ppp}}(\omega_{\text{sfg}}) \propto \left|\mathrm{A}-\mathrm{B}~\text{sin} (3\varphi)\right|^2
\end{equation}
\autoref{eq:ASFG} implies that the $I_{\text{sf}}$ of the Ag(111) surface should have 6-fold symmetry under rotation in $\varphi$, if A $\ll$ B, which is the case for Ag(111)~\cite{Harris:AgASFG:1991}.
As shown previously this underlying structural symmetry is similarly revealed in the measured LEED pattern.
To understand why recall that the Bragg reflections of the LEED pattern result from the intersection of the Ewald sphere with the Ag(111) reciprocal lattice ~\cite{Hove:LEED:1986}. 
The Ag(111) surface lattice is hexagonal, and thus its reciprocal lattice is also hexagonal. 
Thus the LEED pattern of Ag(111) shows a six fold rotational symmetry~\cite{Soares:Agleed:1999,Soares:Agleed:2000}.
\autoref{fig:azimuth}(a) shows the crystal structure, \autoref{fig:azimuth}(b)(top) the azimuthal dependent VSFG signal and \autoref{fig:azimuth}(b)(bottom) the LEED pattern measured for the same sample.  
Clearly the expected 6-fold symmetry is observed, consistent with prior work~\cite{Harris:AgASFG:1991,Friedrich:AgASFG:1993} and the azimuthal dependent nonresonant SFG signal measured within our UHV chamber offers an all-optical probe of surface structural symmetry \cite{Shaw:ASFG:2009, Chen:ASFG:2017, Yang:ASFG:2023, Boyd:ASFG:2020}.

\subsection{Scanning SFG microscopy on Ag(111)}
Having demonstrated that our UHV set-up allows the measurement of azimuthal angle dependent SFG intensities, we finally demonstrate that it allows sample translation in the imaging plane with sufficient stability to image $\approx$ 4~\um sample features. 
As noted above this capability is crucial for the study of TMDC monolayers (whose preparation typically results in $\approx$ 10~\um-sized flakes) thus sample characterization requires optical alignment on the sample. 
A silver sample with two curved lines on surface is prepared through sputtering the same sample position by argon ion for 2 hours and repeating the argon ion sputtering by translating the sample 1~mm in the direction perpendicular to argon ion beam. 
\begin{figure}[!htbp]
	\centering
	\includegraphics[width=0.47\textwidth]{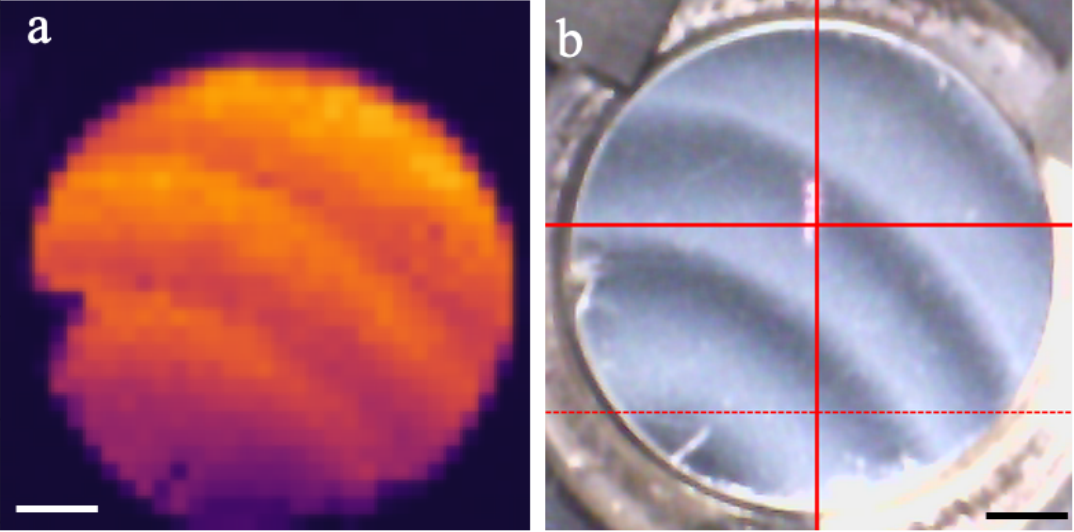}
	\caption{(a) Scanning sum-frequency generation microscopy image of a Ag sample measured by scanning the sample horizontally and vertically. (b) Optical microscopy image of the same area of the same Ag sample. The scale bars represent 1 mm. See text for details.}
	\label{fig:scanning}
\end{figure}

\autoref{fig:scanning}(a) shows a point-wise scanning sum-frequency generation microscopy image of the this Ag sample, obtained by translating the sample, horizontally and vertically, in the image plane.
\autoref{fig:scanning} (b) shows an optical microscopy image of the same area of the same Ag sample. 
Clearly it is possible to image the steps on the Ag surface either by a wide-field visible or scanning SFG microscopy. 
However, because the linear optical susceptibility is a first and second order susceptibility a second rank tensor, the potential information content of the SFG image is higher. 
Point wise scanning or wide-field SFG microscopes for samples under ambient air or electrochemical environments have been reported previously~\cite{Hanninen:SFGM:2017,Zheng:SFGM:2018,Zwaschka:SFGM:2020,Li:SFGM:2020}. 
This measurement is, to our knowledge, the first demonstration of this capability in UHV.

Since SFG is sensitive to surface adsorbates with sub-monolayer sensitivity, scanning SFG microscopy has the potential to probe the -CH, -CO, -OH, etc. vibration and map the spatial distribution of such `fingerprint' vibrations \cite{Hanninen:SFGM:2017}.
While the current set-up is optimized to spectrally resolve features that are $\geq 100~\um$ or larger (as required for experiments involving mechanically exfoliated TMDC samples on Au surfaces), clearly one can imagine systems in which it might be useful to image molecular organization on length scales both shorter (and longer).
As discussed in the methods section increased spatial resolution of down to 21~\um is straightforwardly achievable by a small change in the visible probe beam focussing optics.   
In general terms, we expect 2D heterostructures known to be heterogeneous on a length scale much larger than \emph{single} molecular adsorbates are promising systems to be studied by our scanning SFG microscopy. 
Spatially resolved performance of such heterojunction based devices has been previously described using Optical Beam Induced Current (OBIC)~\cite{Dick:OBIC:1996,Zhuo:OBIC:2022}. 
OBIC utilizes a single color CW optical beam to excite the sample and collects the photocurrent of the heterojunction device by scanning the optical beam in two dimensions to map the uniformity of the local performance.
Such a measurement does not offer correlated structural insight: it's difficult to understand how changes in local structure correlate with device performance. 
Our set-up offers the straightforward possibility of combining OBIC with SFG spectromicroscopy (particularly \textit{operando} while photocurrent is flowing~\cite{huang2023femtosecond}). 
Given this spatial resolution and the symmetry-enabled ability to separate the different contributions to the SFG signal, our set-up seems poised to offer significant new insight into the basic physics of 2D material heterostructures.

\section{Conclusion}
\label{sec:Conclusion}
Probing interfaces between condensed phases and their non-equilibrium dynamics (\textit{e.g.}\ time dependent ultrafast charge and energy transfer), particularly those involving stacks of two dimensional materials, is challenging. 
Because it is largely restricted to linear photon-in/photon-out techniques, much of the difficulty lies in separating the optical response of either bulk phase from those of the 2D material (whether an atomically thin heterostructure or a surface state of a semiconductor).
We and others have previously shown that the symmetry of the second order nonlinear susceptibility, \textit{i.e.}\ the $\chi^{(2)}$, particularly with respect to rotation around the interface normal offers the possibility of near quantitative separation of the optical response of different portions of the interface. 
Such measurements have, virtually exclusively, focussed on heterostructures of solids and atomically thin 2D materials in ambient making temperature control and controlled dosing of molecular condensed matter (\textit{i.e.}\ molecular films) challenging. 

We here describe a set-up that overcomes this limitation: an ultra-high vacuum setup that integrates conventional surface science techniques with azimuthal and polarization dependent, femtosecond time-resolved, sum frequency generation spectroscopy and scanning sum frequency generation microscopy. 
Polarization femtosecond resolved time-averaged vibrationally resonant SFG is demonstrated by its application to a CO monolayer on Pt(111).
Because the set-up allows extraction of \textit{both} the LEED pattern \textit{and} the spectrum under multiple polarizations it is possible to extract the CO interfacial hyperpolarizability ratio (\textit{i.e.} $R = \frac{\beta^{(2)}_{aac}}{\beta^{(2)}_{ccc}}$) without the CO orientations assumed in previous studies. 
We show that the set-up offers femtosecond time-resolution by measuring the so-called free induction decay of the CO stretch vibration on Pt(111), which shows that the spectral line-width measured in the time domain quantitatively reproduces that measured in the frequency domain using a spectrograph/EMICCD camera combination.
The ability to record SFG signals as a function of sample azimuthal rotation is demonstrated for the Ag(111) surface.
Because we measure LEED on the same sample in our UHV system we can directly correlate the 6-fold symmetry of the, nonresonant, SFG response with the 6-fold symmetry of the LEED pattern. 
Finally, we demonstrate sufficient stability to perform scanning SFG microscopy on $\approx$ 10~\um length scales by comparing an optical micrograph and the scanned SFG signal from a, intentionally defect ridden, Ag sample.  

This ability to perform \emph{both} SFG spectroscopy with full characterization of $\chi^{(2)}$ symmetries with femtosecond temporal resolution \emph{and} spatially resolve this signal over 10~\um length scales in UHV appears to offer an important new window onto the physics of two dimensional systems: whether topologically protected surface states, heterostructures of atomically thin two dimensional material or molecular-adsorbate solid interfaces.

\begin{acknowledgments}\noindent
We want to thank H. Kirsch and S. Kubala for valuable discussions regarding the initial design of the setup. We thank M. Lackner for his help in developing the control software for the setup and E. Hasselbrink for donating his LEED/AES device. This work was supported by the Deutsche Forschungsgemeinschaft (DFG, German Research Foundation) through Projects A06 of the Collaborative Research Center SFB 1242 “Non-Equilibrium Dynamics of Condensed Matter in the Time Domain” (Project No. 278162697), through Germany’s Excellence Strategy (EXC 2033 - 390677874 - RESOLV) and through SCHL 384/20 - 1 (Project No. 406129719). Additional support was provided by the European Research Council, i.e., ERC - CoG - 2017 SOLWET (Project No. 772286) to R.K.C. 
\end{acknowledgments}

\end{document}